\documentclass[aps,prl,reprint,superscriptaddress,floatfix]{revtex4-2}
\usepackage{amsmath,bm,mathtools,graphicx}
\usepackage{multirow, booktabs}

\begin{document}

\title{Altermagnetic pseudogap from $\frac{t}{u}$ expansion}
\begin{abstract}
    Order parameter analysis of the t/U series reveals a uniform altermagnet endemic to the doped Mott insulator, driven by kinetic interactions, occupying a position between the antiferromagnet and hole-doped d-wave superconductor that is normally reserved for the pseudogap. The metastable boundary of the altermagnet punctures and divides the superconductor into underdoped and overdoped regions, reminiscent of the $T^*$ crossover or transition in the cuprates. Similarly, the $T_{pair}$ boundary of the superconductor divides the altermagnet, leading to a low temperature phase susceptible to Cooper fluctuations. The altermagnet is unstable to inhomegeneous spin and charge order of sites, bonds, and currents. Its leading instability is to the $\pi$-flux state, suggesting the possible emergence of spin-charge liquids and  quantum ordered states from a physically realistic microscopic model.
\end{abstract}
\author{Rohit Hegde}
\email{rohegde@gmail.com}
\noaffiliation
\date{\today}

\maketitle
\emph{Introduction}.---Forty years of research on the cuprates have brought a variety of exotic actors to the stage beyond high-temperature superconductivity. The pseudogap has long registered as mysterious and paradoxical, as a potentially ordered phase lacking a clear order parameter. An enduring question is whether the order is symmetry-breaking or long-range entangled, while a fundamental challenge to modeling is the fact that empirically the pseudogap is not a monolith, but rather a collection of experimental signatures—such as a partial depletion of low-energy spectral weight, Fermi arcs, and a transport-defined crossover scale $T^\ast$—emerging over a range of temperatures and dopings across different materials \cite{timusk_pseudogap_1999,norman_pseudogap_2005,keimer_quantum_2015,ding_spectroscopic_1996,ando_electronic_2004}.

Theoretical interpretations of the pseudogap have ranged from precursor superconductivity and phase-fluctuating pairing above $T_c$ \cite{kanigel_evidence_2008,kanigel_protected_2007,emery_phase_1995} , to competing or intertwined orders such as density waves and pair-density waves \cite{agterberg_physics_2020,berg_dynamical_2007,chakravarty_hidden_2001} , as well as more exotic fractionalized gauge-theory descriptions of the doped Mott insulator \cite{wenlee_underdoped_1996,lee_nagaosa_wen_2006,read_sachdev_valence_1989}. A persistent theme is that the pseudogap occupies the region of the phase diagram between antiferromagnetism and $d$-wave superconductivity, suggesting that the underlying physics is closely tied to the doped Mott insulator and proximity to associated quantum phase transitions \cite{keimer_quantum_2015,arovas_hubbard_2022,sachdev_order_2003}.

In parallel, recent work has identified altermagnetism as a distinct symmetry class of collinear spin order characterized by vanishing net magnetization together with momentum-dependent spin splitting \cite{smejkal_emerging_2022,jungwirth_altermagnetism_2025}. Interest in altermagnetic states has grown rapidly, including studies of quasiparticles and collective modes \cite{rostami_fermi_2025}. In microscopic models, interaction-driven $d$-wave altermagnetism can emerge spontaneously near antiferromagnetic regimes, including constructions based on coexisting AFM and orbital order or AFM with $d$-wave spin-bond order \cite{leeb_spontaneous_2024,dong_spontaneous_2025,li_enhancement_2025,das_realizing_2024}.

In this work we show that a uniform $d$-wave altermagnet emerges naturally within the $t/U$ expansion of the Hubbard model \cite{macdonald_t_1988}. Mean-field theory of the Hubbard-commuting dynamic reveals a state driven entirely by kinetic interactions occupying the portion of the phase diagram between antiferromagnetism and hole-doped $d$-wave superconductivity commonly associated with the pseudogap. The metastable boundary of the altermagnet punctures the superconducting dome in a manner reminiscent of the $T^\ast$ line, while pairing fluctuations subdivide the altermagnetic regime into regions more and less susceptible to Cooper pairing. Beyond the uniform state, analysis of collective fluctuations reveals strong instabilities toward bond currents and $\pi$-flux patterns, suggesting a route from a simple altermagnetic parent state to the more complex intertwined orders often discussed in the pseudogap regime.

\begin{figure}
    \centering
    \includegraphics[scale=0.076]{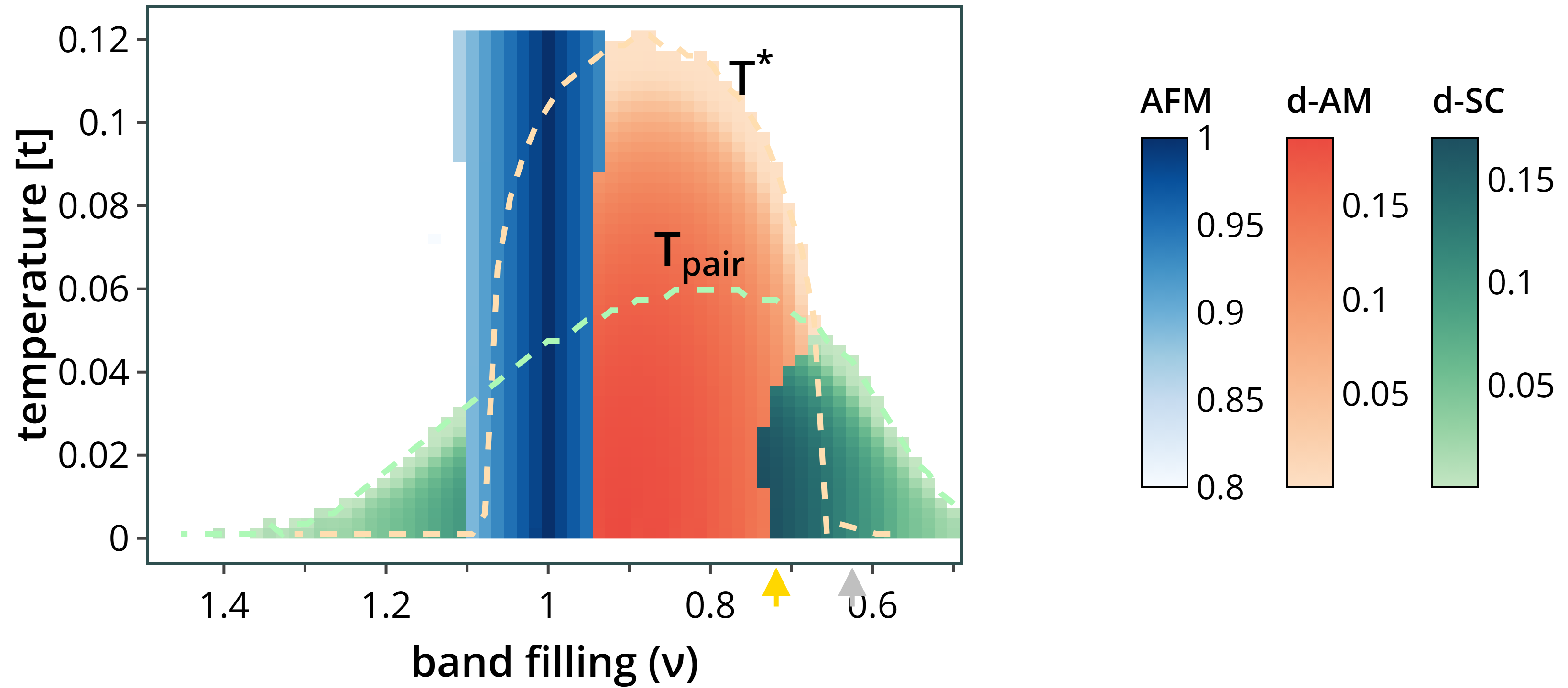}
    \caption{Free energy phase diagram from static mean-field theory of t/U model \eqref{tUmodel} including antiferromagnetism, q=0 altermagnetism, superconductivity, and the Fermi gas (white). Altermagnetism here is supported by a kinetic interaction that is deactivated by the purely localized Mott insulator, so is more like a Pomeranchuk instability of a Fermi liquid than a consequence of antiferromagnetism, but all orders are siblings that arise together from a Hubbard-commuting dynamic. Dynamical parameters $U, J$ are effectively renormalized to compensate for lack of correlations,  $U/|t|=0.80, J/|t|=1.1, t'/t=-1/7$. Metastable phase boundaries of d-AM and d-SC are presented as $T^*$ and $T_{pair}$. $T^*$ cleaves the hole-doped superconductor into underdoped and overdoped sides, while in the altermagnetic phase, $T_{pair}$ signals onset of strong Cooper pair fluctuations. The altermagnetic phase is fully concealed by AFM for $\nu>1$.   }
    \label{fig:phasediagram}
\end{figure}

\emph{Model of electrons}.---Hubbard-commutativity is the defining symmetry of the t/U series expansion at finite order. At first order is a kinetic energy of doublons and holons, free of creation and annihilation terms that change their number, as in the model Hamiltonian,
\begin{equation}
    H=U+T_h+T_d+J_s+J_c \, .
    \label{tUmodel}
\end{equation}
Commutativity allows for kinetics of the form $\alpha T_h + \beta T_d$, resulting in a generalized and expansive sense of charge asymmetry \cite{hegde_charge_2024}, but the choice $\alpha=\beta=1$ as in the t/U series renders an antisymmetric operation, just like the non-interacting bandstructure. The implication for the standard two-parameter ($t,t'$) square-lattice hopping model deployed here is that sign and magnitude of $t'/t$ impart asymmetry to the overall dynamic as in the more familiar Hubbard model. Crucially, when expressed in terms of electronic fields, the kinetic $T_h+T_d=T-T_2+2T_3$ contains a three-body part, $T_3=\sum_{ij,\sigma}c^\dagger_{i\sigma}c_{j\sigma}n_{i-\sigma}n_{j-\sigma}$, as the herald of asymmetry, in analogy to the three-electron Haldane pseudopotential interaction that differentiates the Pfaffian and anti-Pfaffian states of the fractional quantum Hall regime. In the Hubbard setting, $T_3$ is decisive in favoring some orders while disfavoring others, and is directly responsible for the appearance of altermagnetism. When Wick symmetrized and expressed in terms of bilinears,
\begin{align}
    T_3 = \frac{1}{4}\sum_{ij}t_{ij}\Big[ \rho_{ij} \big( \rho_i \rho_j +\bm{\sigma}_i \bm{\sigma}_j - \rho_{ji}^2 + \bm{\sigma}_{ji}^2 \big) \nonumber \\
    -\bm{\sigma}_{ij} (\rho_i \bm{\sigma}_j+\rho_j \bm{\sigma}_i) + i\epsilon_{abc}\sigma^a_{ij} \sigma^b_i \sigma^c_j         \Big] \,. \label{T3}
\end{align}
Other kinetic operations present are the bare band hopping, $T=\sum_{ij}t_{ij}\rho_{ij}$, and correlated hopping, $T_2 = (1/2)\sum_{ij}t_{ij}(\rho_{ij}(\rho_i+\rho_j)-\bm{\sigma}_{ij}(\bm{\sigma}_i+\bm{\sigma}_j))$.

The terms at second order are technically complicated, ranging from two to five body operations valued over two or three lattice sites. First neighbor spin exchange, $J_s=(J/4)\sum_{(ij)}(\bm{\sigma}_i\bm{\sigma}_j-\rho_i\rho_j)$, is most familiar and indispensable. This letter highlights the significant impact of charge exchange (pair-hopping), $J_c=(J/2)\sum_{(ij)}( c^\dagger_{i\uparrow}c^\dagger_{i\downarrow} c_{j\downarrow}c_{j\uparrow}+c^\dagger_{j\uparrow}c^\dagger_{j\downarrow} c_{i\downarrow}c_{i\uparrow})$, as shown in Table \ref{tab:order_symmetries}. In the particle-hole channel, $J_c$ couples to bond orders in a way that cancels $J_s$ for inversion-symmetric hoppings, but doubly reinforces it for currents.

\begin{table}[t]
\centering
\caption{Symmetry properties and dynamical exchange couplings for nearest-neighbor-bond orders. $\mathcal{P}$ and $\mathcal{T}$ are inversion and time-reversal. Spin-charge refers to the coupling of the effective two-body interaction generated from $T_3$. For example, the total effective interaction of spin currents is repulsive, $\sum_{(ij)}(J/2-t \langle \rho_{ij} \rangle) |j^z_{ij}|^2$. Since $t \langle \rho_{ij} \rangle <0$, spin hopping is the only symmetry-breaking condensate that lowers the kinetic energy. When spin-spin and charge-charge exchange are counted together, the $\mathcal{O}(t^2/u)$ dynamic decouples from charge and spin hopping.}
\label{tab:order_symmetries}
\begin{tabular*}{\columnwidth}{@{\extracolsep{\fill}} l cc ccc}
\toprule
\multirow{2}{*}{Bond Order} & \multicolumn{2}{c}{Symmetry} & \multicolumn{3}{c}{Exchange Coupling} \\
\cmidrule(lr){2-3} \cmidrule(lr){4-6}
 & $\mathcal{P}$ & $\mathcal{T}$ & sp-ch & sp--sp & ch--ch \\
\midrule
Charge Hopping & $+$ & $+$ & $-3 t \langle \rho_{ij} \rangle $ & $-J/4$ & $J/4$ \\
Charge Current & $-$ & $-$ & $-t \langle \rho_{ij} \rangle $ & $-J/4$ & $-J/4$ \\
Spin Hopping   & $+$ & $-$ & $+t \langle \rho_{ij} \rangle$ & $J/4$ & $-J/4$ \\
Spin Current   & $-$ & $+$ & $-t \langle \rho_{ij} \rangle$ & $J/4$ & $J/4$ \\
Cooper Singlet & $+$ & $+$ & $-t \langle \rho_{ij} \rangle$ & $-J/2$ & $0$ \\
\bottomrule
\end{tabular*}
\end{table}
\textit{Mean-field phases}.---In an attempt to uncover the essential imprint of the dynamic \eqref{tUmodel} on the equilibrium order phase diagram, we consider a highly restricted mean-field that includes the combination of antiferromagnetism (AFM), d-wave spin-hopping-type altermagnetism (d-AM), d-density-wave order (d-DW), and uniform superconductivity (d-SC). This list notably excludes breaking of rotational symmetry or translation symmetry (beyond the doubled unit cell). 

The quasiparticle Hamiltonian is diagonal in magnetic zone momentum, and operates in an eight-component space $\sigma \otimes \tau\otimes\gamma$, referencing spin, sublattice, and Nambu charge,
\begin{align}
    H_{qp}(\bm{k})=4 t \big(c_s s(\bm{k}) +c_a d(\bm{k}) \sigma^z\big) \tau^x \gamma^z + c_j d(\bm{k})\tau^y  \gamma^z \nonumber \\
    +c_m \sigma^z \tau^z \gamma^z  +4 t' s'(\bm{k}) \big(c_{s'} + c_{ms'} \sigma^z\tau^z \big)\gamma^z \nonumber \\
    +(c_{\Delta} \gamma^- - c_\Delta^* \gamma^+) d(\bm{k}) (-i \sigma^y) \tau^x -\tilde{\mu} \gamma^z ,
    \label{quasiparticle}
\end{align}
where $s(\bm{k})/d(\bm{k}) = (\cos{k_x}\pm \cos{k_y})/2$, $s'(\bm{k})=\cos{k_x}\cos{k_y}$ are unit norm form factors. Mean-fields are defined such that $c_s, c_a, c_{s'}, c_{ms'} $ are dimensionless while $c_j, c_m, c_\Delta, \tilde{\mu}$ have units of energy. The fields $c_s, c_{s'}$ renormalize the bare hoppings $t,t'$ and depend not only on the filling $\nu$ but on the bond kinetic energies and the rest of the order parameters. Expression of the self-consistent fields is presented in SM. Although the Hamiltonian \eqref{quasiparticle} allows for combinations of orders, it is found that the phase diagram in Fig \ref{fig:phasediagram} contains only singular condensates.

The essence of the AFM at $\nu=1$ is localization, not only in the sense of the wavefunction, $\langle \mathcal{O}_{ij}\rangle_{AFM} \propto \delta_{ij}$ for all bilinears $\mathcal{O}$, but in that chiral symmetry forces the quasiparticles to disperse only with $t'$ and not with $t$, as in $c_s=0, c_{s'}=1$, narrowing their bandwidth so the fully polarized $\langle m\rangle_{AFM}=1$ state is metastable if $U+2J \ge 8t'$. The AFM remains nearly fully polarized away from $\nu=1$, so largely avoids the energetic cost of the Hubbard interaction everywhere it appears, and completely avoids it in the emergent Heisenberg model manifesting at $\nu=1, T=0K$. The AFM showcases a sharp discrepancy between static order and time-dependent fluctuations. In the static theory, as in Fig \ref{fig:phasediagram}, the value of $1-\langle m \rangle_{AFM,HF} < 10^{-4}$ up to the highest temperatures at half-filling, while in linear spin wave theory, the demagnetization is $1-\langle m \rangle_{AFM,TDHF} =0.394$ at zero temperature.

\begin{figure}[t]
    \centering
    \includegraphics[scale=0.076]{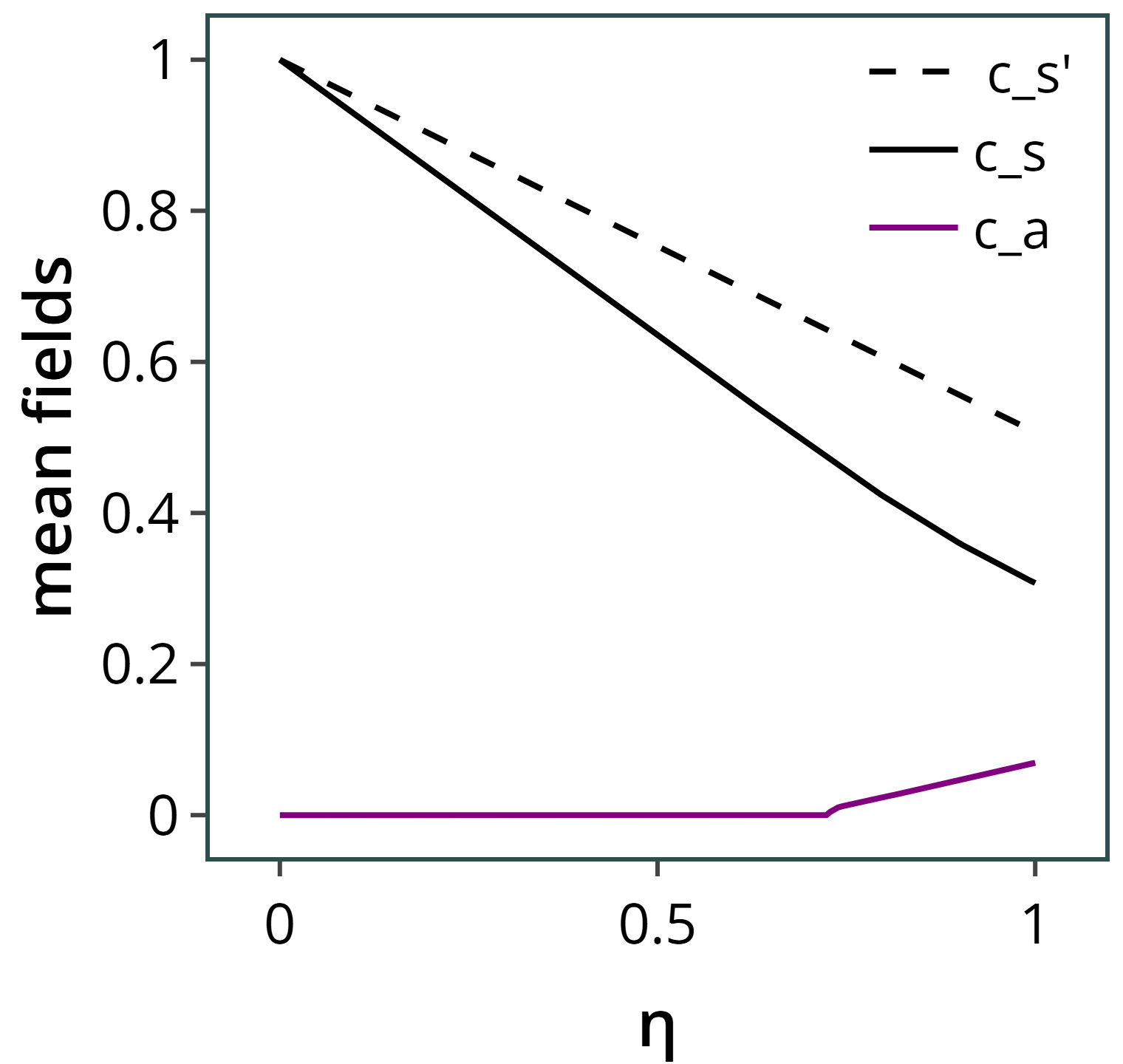}
    \includegraphics[scale=0.076]{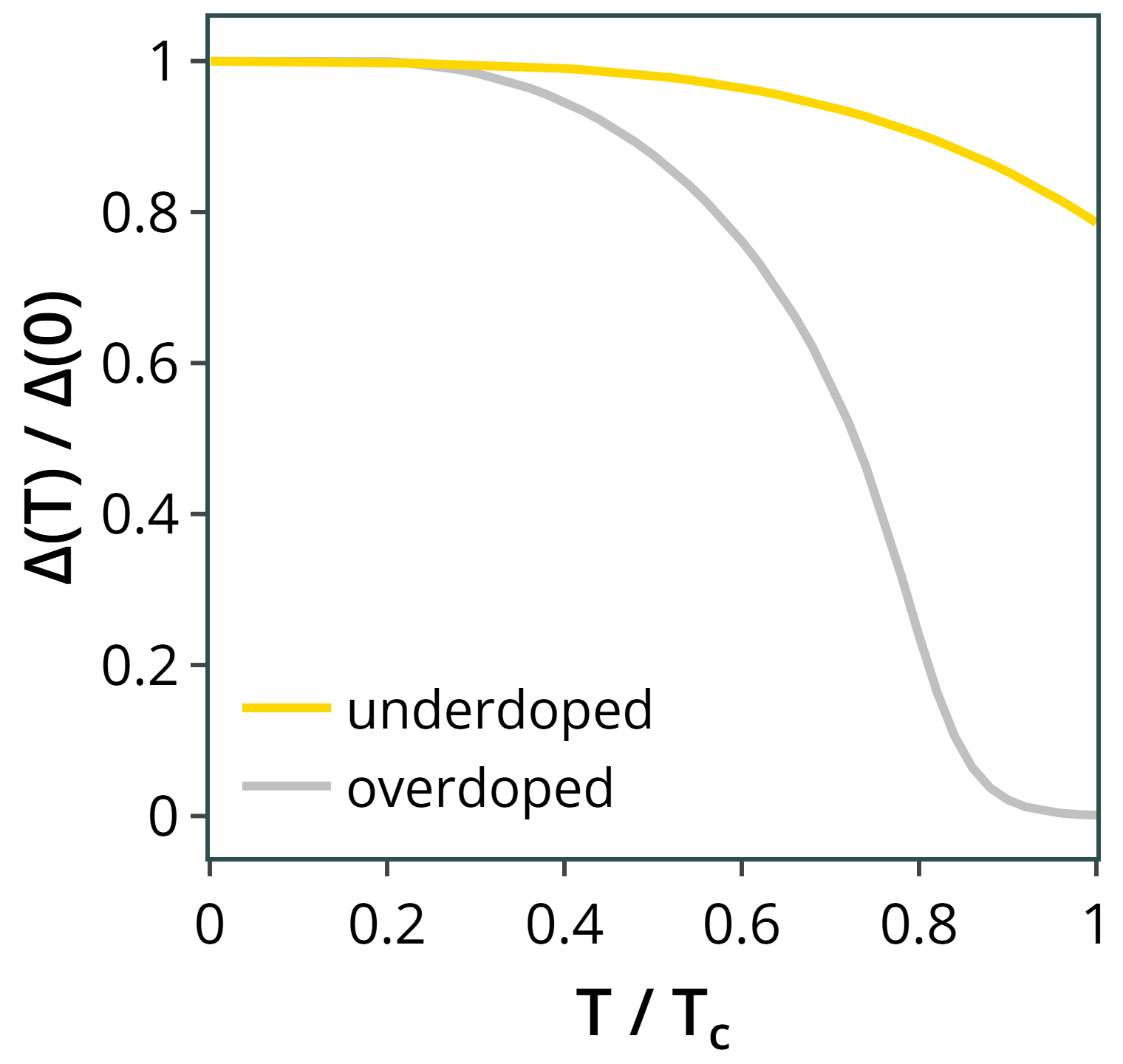}
    \caption{\textbf{a}) Tuning from non-interacting band to Hubbard-commuting kinetic, $T(\eta)=T_h+T_d+(1-\eta)(T_++T_-)$, first and second neighbor mean-field hoppings diminish in scale, $c_s, c_{s'}$, as non-commuting processes are removed. When $\eta \sim1$, altermagnetism condenses in energetic sympathy with the s-wave band. \textbf{b}) The overdoped superconductor is like standard BCS in that the pairing gap vanishes at $T_c$. The underdoped d-SC is distinguished by its melting transition to another ordered phase. A potentially higher $T_c$ could be unlocked by the targeted suppression of altermagnetism.  }
    \label{fig:phasetransition}
\end{figure}

Altermagnetism, by contrast, is an electronic order that fosters delocalization of doublons and holons. Its sole dynamical driver is the piece of $2 T_3$ that is $(1/2)\sum_{(ij)}t(\rho_{ij}\bm{\sigma}_{ji}^2+\rho_{ji}\bm{\sigma}_{ij}^2) $, implying for the quasiparticles that $c_a=\langle a\rangle \langle \rho_{i,i+x}\rangle$, and $c_s=\ldots +|\langle a \rangle|^2 /2+\ldots$, meaning that d-AM requires s-wave delocalization to condense, and that the AM condensate in turn enhances the effective s-wave hopping $c_s t$. One consequence is the incompatibility of altermagnetism and pure Mott insulation, since $c_a=0$ in the MI phase. A quantum phase transition between Fermi gas and altermagnet is uncovered in linear interpolation between bare hopping ($\eta=0$) and Hubbard-commuting ($\eta=1$), $T(\eta)=T+\eta(2T_3-T_2)$, depicted in Fig \ref{fig:phasetransition}. As $\eta$ increases after altermagnetism condenses, s-wave hopping $c_s$ curves slightly upward, countering the basic trend towards localization. Within mean-field theory, the AM has the same Hubbard energy as the Fermi gas, $U\nu^2/4$, since the bond-spin condensate is not counted by the on-site interaction. Given this distinction between AFM and AM, U can be tuned to set the phase boundary between them. If U is set to bandwidth scale in the static theory, AFM will dominate over much of the phase diagram, leaving little room for other phases. In Fig \ref{fig:phasediagram}, the value of $U=0.8$ was chosen in conjunction with $J$ to set $\nu_{c,AFM-AM} \sim 0.95$. 

In the model \eqref{tUmodel}, d-wave superconductivity is only supported by spin exchange $J_s$, is suppressed by $T_3$, and decouples from $T_2$ and $U$. By contrast, first neighbor s-wave pairing is additionally suppressed by $T_2$ and is mixed by it into on-site s-pairing, which is directly penalized by $U$. Unlike in renormalized mean-field theories of the t-J model, in the mean-field theory \eqref{quasiparticle} s- and d-wave superconductivity are not energetically competitive, but are largely split by the interactions that make Hubbard-commuting symmetry. Tuning the value of $J$ forms the phase boundary between hole-doped superconductor and altermagnet. Given that $c_\Delta=-(2J+4t\langle \rho_{i,i+x}\rangle )\Delta^* $, and first neighbor bond order amounts to $\rho_{i,i+x}\sim0.4$ in the Fermi gas, $J>0.8$ to support Cooper pairing. This is another example of bias in the absence of correlations, which if present would reduce measurements of bond order in delocalized phases like the FG and AM. In MFT, bond order can be suppressed to zero in local product states like the classical Mott insulators, but for k-space products it tends to the Fermi gas value, while the primary mechanism for restricting kinetic energy is the suppression of effective couplings. In a fully quantum theory, couplings would be unaltered, and localization would manifest in measurement. In Fig \ref{fig:phasediagram}, the value of $J=1.1$ allows for regions of robust d-SC on both sides of $\nu=1$, and also makes d-SC competitive with AM, leading to emergence of the underdoped d-SC. Adding the metastable phase boundaries for AM and d-SC, called $T^*$ and $T_{pair}$ makes clear that the underdoped region is where d-SC outcompetes AM. The metastable boundary of a phase is its border with the Fermi gas, representing the maximum extent of its support. Optimal $T_c$ of the d-SC emerges as the intersection of $T^*$ and $T_{pair}$, and the mean-field character of the underdoped side is distinguished by non-vanishing of the order parameter at $T_c$, as illustrated in Fig \ref{fig:phasetransition}b. Filling factors corresponding to underdoped/overdoped are marked by gold/silver arrows in Fig \ref{fig:phasediagram}. The fact that much of the altermagnet sits below $T_{pair}$ naturally explains enchanced pairing susceptibility without requiring pre-formed Cooper pairs. The D-density wave state shares the same interaction coupling as d-SC, and is supported in MFT as a metastable state, but is outcompeted by d-SC. When the mean-field algorithm is simultaneously seeded with DDW and d-SC orders, the former decays to zero, while the latter survives as condensate. 

\begin{figure}
    \centering
    \includegraphics[scale=0.076]{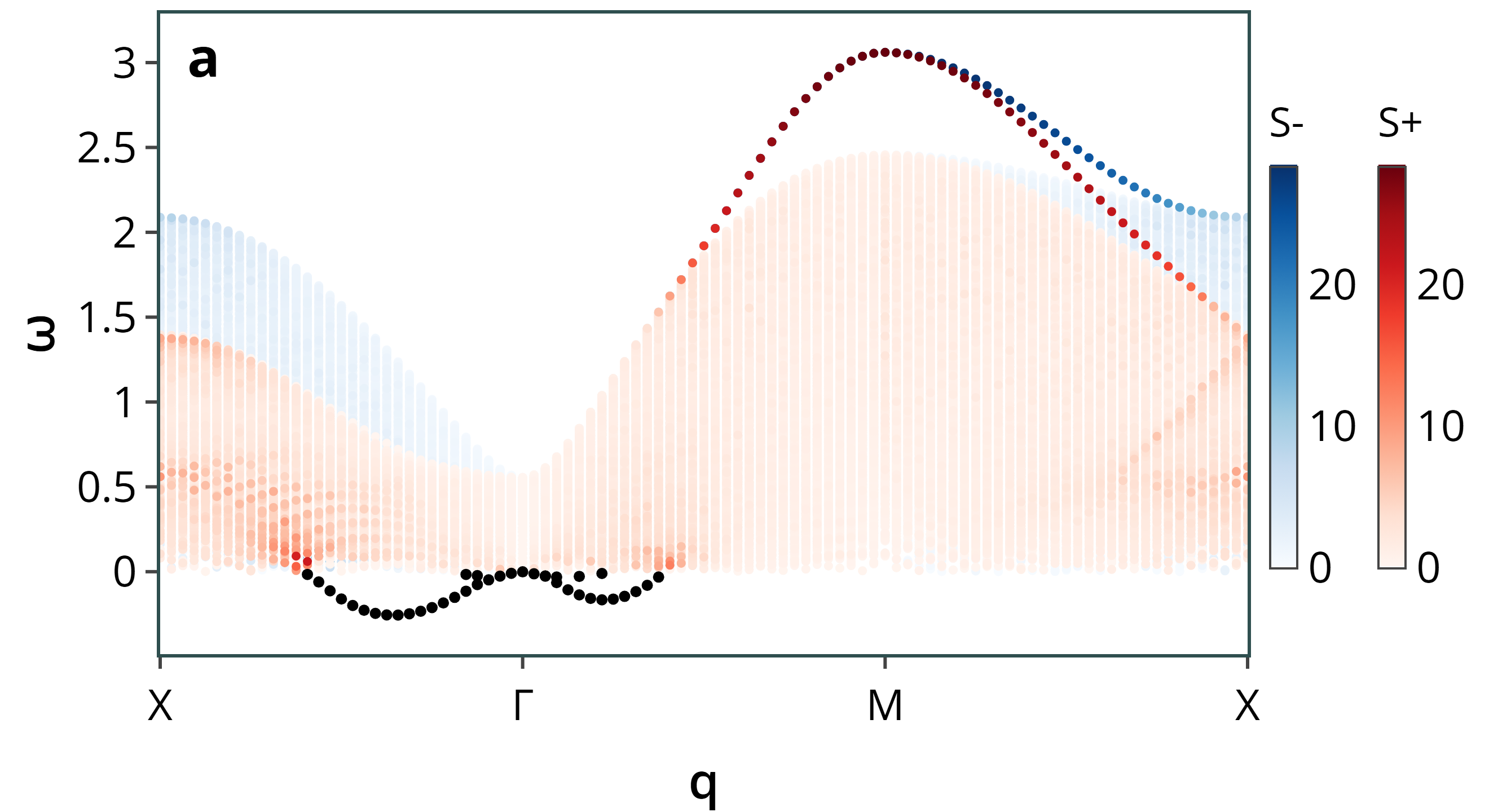}
    \includegraphics[scale=0.076]{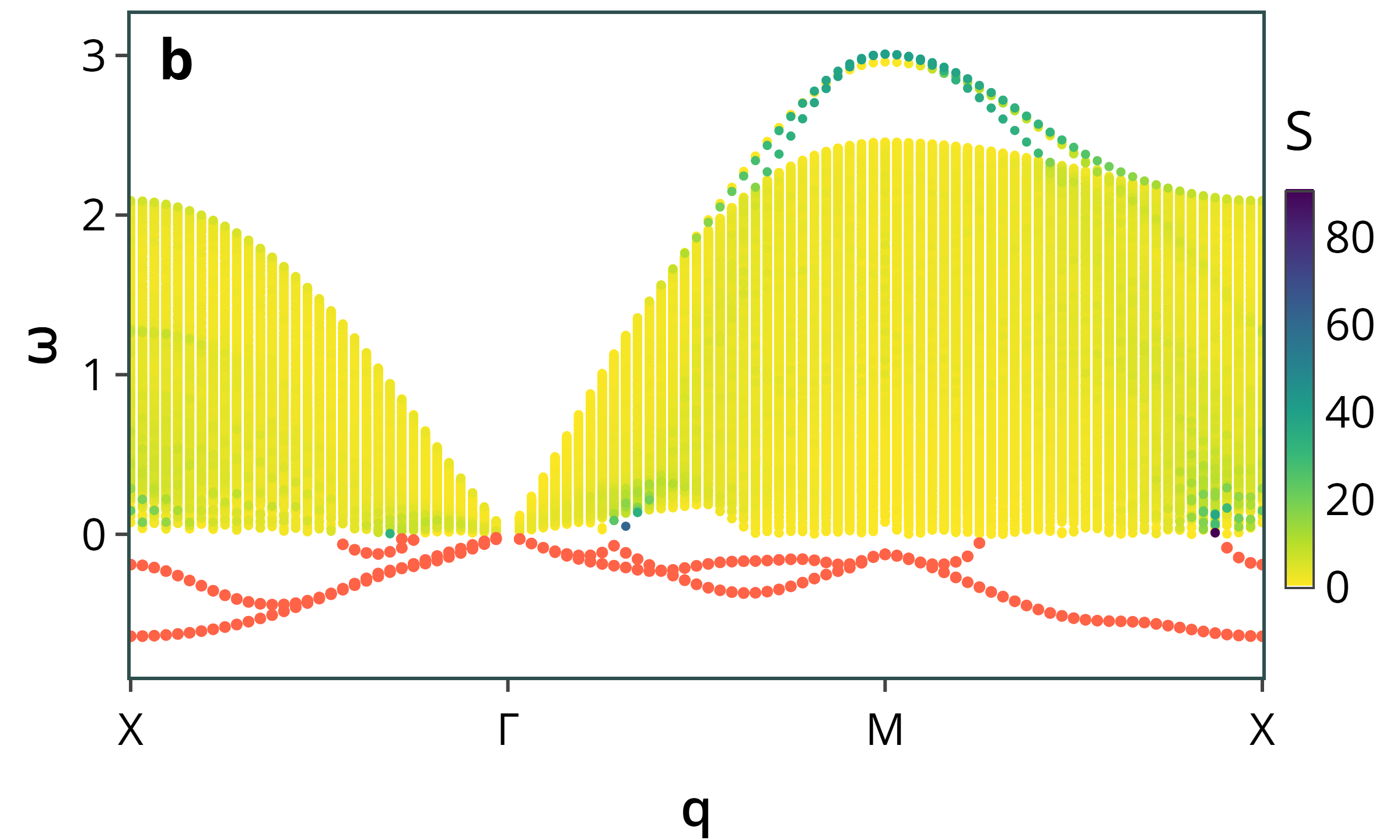}
    \includegraphics[scale=0.076]{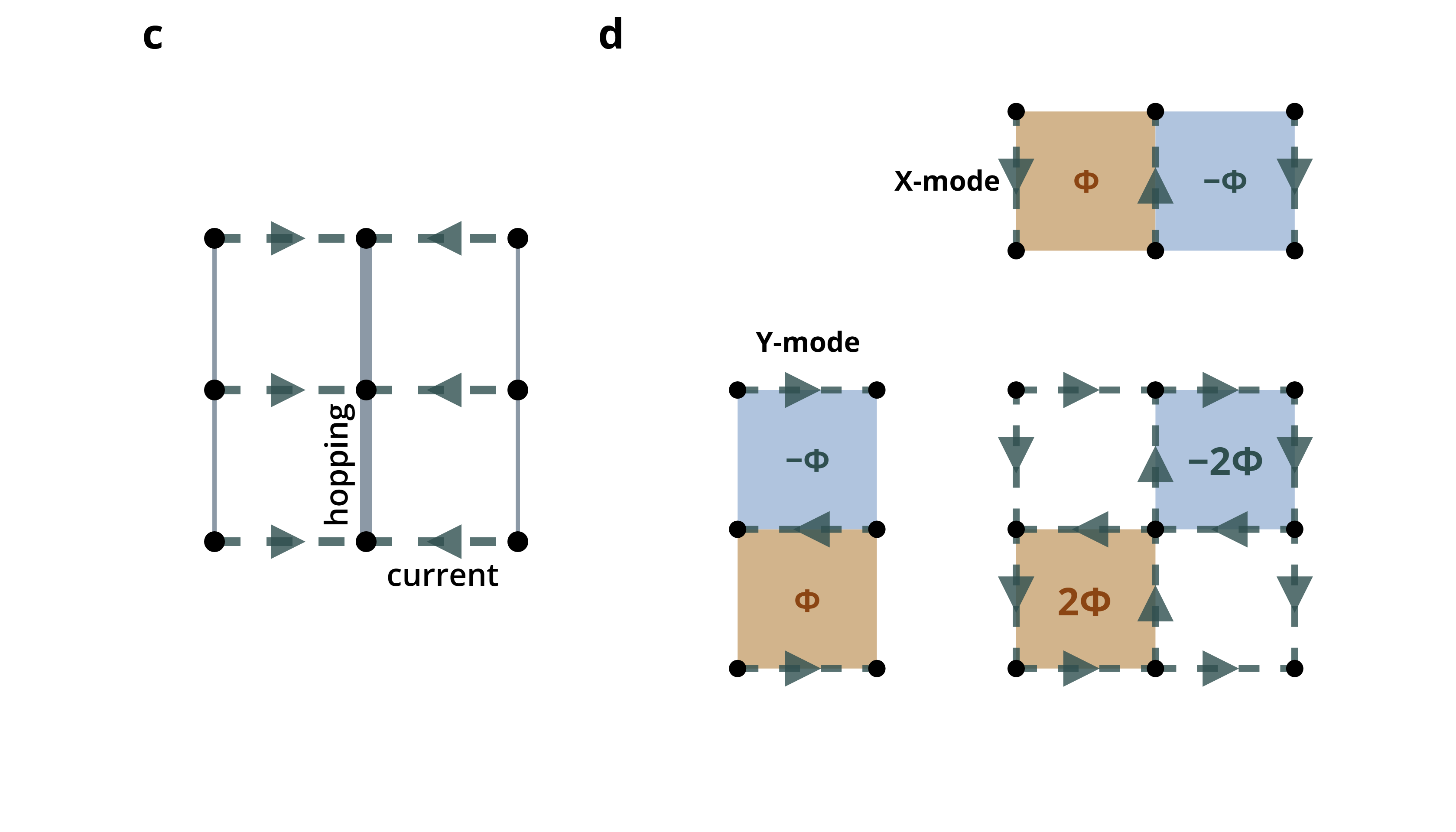}
    \caption{\textbf{a)} Transverse susceptibility spectrum of the altermagnet within TDHF theory, for a simplified Hamiltonian with $U=J=0$. A purely d-wave Goldstone boson becomes unstable and anisotropic away from $\bm{q}=\Gamma$, and disperses mostly as an s-wave paramagnon near $\bm{q}=M$. \textbf{b)} The most severe instability is common to the AM and the Fermi gas, appearing in the charge sector at the antinodes $\bm{q}=X,Y$ nucleating $\pi$-flux order. Superposition of $X$ and $Y$ modes on top of background s-wave hopping leads to a checkerboard of fluxes, as in \textbf{d)}. More particuar to the AM is a secondary instability at $\bm{q}=X,Y$ intertwining longitudinal currents with transverse modulation of hopping, as in \textbf{c)}.  }
    \label{fig:fluctuation}
\end{figure}

\emph{Fluctuating altermagnet}.---In analyzing fluctuations, we focus on the kinetic interactions that make altermagnetism, and set $U=J=0$ in the time-dependent Hartree-Fock calculations presented in Fig \ref{fig:fluctuation}. Uniform altermagnetism only breaks spin symmetry on first neighbor bonds in the d-wave channel. Correspondingly, a Nambu-Goldstone boson appears in the transverse susceptibility spectrum, with an internal wavefunction that is purely d-wave. The mode is unstable away from $\bm{q}=\Gamma$, maximally so at $q=(3\pi/8a) \bm{e_x}$ or $q=(3\pi/8a) \bm{e_y}$, where the wavefunction of p-h pairs mostly represents on-site spin fluctuation, and secondarily contains a purely anisotropic spin current in the lattice direction parallel to the mode wavevector. A nearly identical instability appears in the Fermi gas, so can be understood as fundamental to the dynamic, rather than to the symmtery broken state. Indeed, away from $\bm{q}=\Gamma$, the collective excitation appears mostly para-magnonic, in that it expresses largely in the on-site s-wave channel in which symmetry is not broken. The transverse spectrum has the expected property of an altermagnet, $\chi^+(q_x,q_y,\omega)=\chi^-(q_y,-q_x,\omega)$.

Instabilites are more rampant in the charge sector, as illustrated in Fig \ref{fig:fluctuation}b. The most severe instability is at $\bm{q}=X,Y$, appearing as alternating charge currents flowing on bonds perpendicular to the mode wavevector. This pattern corresponds to an electronic version of the well-known $\pi$-flux spin-liquid state, as depicted in Fig \ref{fig:fluctuation}d. The $\pi$-flux state is understood to be degenerate with other orders in the chargeless sector at half-filling, but is distinct and non-degenerate elsewhere. Superposing the instabilities at $X$ and $Y$ results in a fluxless state, but if the instabilities coalesce on top of the regular and expectedly present s-wave bond kinetic energy, the result is a checkerboard of plaquettes threaded by fluxes of $0, 2\Phi,-2\Phi$. The flux modes at $X$ and $Y$ are also the primary instabilities of the Fermi gas, when driven by the Hamiltonian $T-T_2+2T_3$. More specific to the altermagnet is the emergence of a secondary charge instability at the anti-nodes, which mixes longitudinal currents with transverse modulation of hopping, as in Fig \ref{fig:fluctuation}c.

\emph{Discussion.}---The emergence of uniform $d$-wave altermagnetism within the $t/U$ expansion offers a simply parameterized parent state for the pseudogap regime. Its inherent instability toward complex, intertwined spin and charge modulations—most notably $\pi$-flux patterns and bond currents—suggests that the experimentally observed macroscopic heterogeneity of the pseudogap may naturally cascade from this precursor. A particularly rich avenue for future exploration is the possibility of an AM* phase, in which portions of the altermagnetic condensate survive while other degrees of freedom strongly fluctuate and fractionalize.

Building on this inherent instability toward charge order, a critical next step is to investigate the emergence of pair density waves (PDWs) from Hubbard-commuting dynamics. A defining feature of the PDWs observed in cuprates—distinguishing them from better-understood FFLO states—is their lattice-commensurate, short-ranged periodicity (e.g., $4a_0$ or $8a_0$). Such periods emerge naturally from a kinetic framework where doublons and holons delocalize by exchanging places with singlons. This mechanism motivates the definition of a far-intermediate-coupling regime: one where dynamics are effectively Hubbard-commuting, but the on-site interaction $U$ remains smaller than the bandwidth, sustaining a higher density of doublons and holons than the strict strong-coupling limit. Crucially, in this regime, the on-site interaction is weaker than the effective kinetic interactions on short-distance bonds. This resulting spatial non-monotonicity in the interaction profile has recently been identified as a general driver for PDW formation \cite{wu_pair_2023}.

While our static mean-field boundaries capture vital phenomenological features like the $T^\ast$ and $T_{pair}$ crossovers, severe collective charge instabilities point to a deeper connection to quantum spin liquid physics. The shared $\pi$-flux instability between the altermagnet and the Fermi gas highlights a fundamental consequence of the Hubbard-commuting dynamic. Future work deploying partonic or ancilla constructions on the $t/U$ model could illuminate how order melts into quantum order, bridging the gap between classical symmetry breaking and long-range entanglement in strongly correlated systems.

\begin{acknowledgments}
\emph{Acknowledgments.}---I wish to thank Yafis Barlas, Gaurav Chaudhary, Hua Chen, Maine Christos, Eslam Khalaf, Bishoy Kousa, Patrick Ledwith, Allan MacDonald, Jose Manuel Torres López, Sparsh Mishra, Nicolás Morales-Durán, Dima Pesin, Rhine Samajdar, Jingtian Shi, Inti Sodemann, Tobias Wolf, and Fan Zhang for helpful discussions.
\end{acknowledgments}

\nocite{*}
\bibliography{main}

@misc{yang_emergent_2026,
	title = {Emergent spin-resolved electronic charge density waves and pseudogap phenomena from strong $d$-wave altermagnetism},
	author = {Yang, Fei and Zhao, Guo-Dong and Yan, Binghai and Chen, Long-Qing},
	year = {2026},
	howpublished = {arXiv:2602.11694 [cond-mat.str-el]},
	doi = {10.48550/arXiv.2602.11694}
}

@article{agterberg_physics_2020,
	title = {The Physics of Pair Density Waves},
	author = {Agterberg, Daniel F. and Davis, J. C. S{\'e}amus and Edkins, Stephen D. and Fradkin, Eduardo and Van Harlingen, Dale J. and Kivelson, Steven A. and Lee, Patrick A. and Radzihovsky, Leo and Tranquada, John M. and Wang, Yuxuan},
	journal = {Annu. Rev. Condens. Matter Phys.},
	volume = {11},
	pages = {231},
	year = {2020},
	doi = {10.1146/annurev-conmatphys-031119-050711}
}

@article{lee_amperean_2014,
	title = {Amperean Pairing and the Pseudogap Phase of Cuprate Superconductors},
	author = {Lee, Patrick A.},
	journal = {Phys. Rev. X},
	volume = {4},
	pages = {031017},
	year = {2014},
	doi = {10.1103/PhysRevX.4.031017}
}

@article{berg_dynamical_2007,
	title = {Dynamical Layer Decoupling in a Stripe-Ordered High-$T_c$ Superconductor},
	author = {Berg, E. and Fradkin, E. and Kim, E.-A. and Kivelson, S. A. and Oganesyan, V. and Tranquada, J. M. and Zhang, S. C.},
	journal = {Phys. Rev. Lett.},
	volume = {99},
	pages = {127003},
	year = {2007},
	doi = {10.1103/PhysRevLett.99.127003}
}

@article{hashimoto_energy_2014,
	title = {Energy gaps in high-transition-temperature cuprate superconductors},
	author = {Hashimoto, Makoto and Vishik, Inna M. and He, Rui-Hua and Devereaux, Thomas P. and Shen, Zhi-Xun},
	journal = {Nature Phys.},
	volume = {10},
	pages = {483},
	year = {2014},
	doi = {10.1038/nphys3009}
}

@article{keimer_quantum_2015,
	title = {From quantum matter to high-temperature superconductivity in copper oxides},
	author = {Keimer, B. and Kivelson, S. A. and Norman, M. R. and Uchida, S. and Zaanen, J.},
	journal = {Nature},
	volume = {518},
	pages = {179},
	year = {2015},
	doi = {10.1038/nature14165}
}

@article{jungwirth_altermagnetism_2025,
	title = {Altermagnetism: {An} unconventional spin-ordered phase of matter},
	author = {Jungwirth, Tom{\'a}{\v s} and Fernandes, Rafael M. and Fradkin, Eduardo and MacDonald, Allan H. and Sinova, Jairo and {\v S}mejkal, Libor},
	journal = {Newton},
	volume = {1},
	pages = {100162},
	year = {2025},
	doi = {10.1016/j.newton.2025.100162}
}

@article{kanigel_evidence_2008,
	title = {Evidence for Pairing above the Transition Temperature of Cuprate Superconductors from the Electronic Dispersion in the Pseudogap Phase},
	author = {Kanigel, A. and Chatterjee, U. and Randeria, M. and Norman, M. R. and Koren, G. and Kadowaki, K. and Campuzano, J. C.},
	journal = {Phys. Rev. Lett.},
	volume = {101},
	pages = {137002},
	year = {2008},
	doi = {10.1103/PhysRevLett.101.137002}
}

@article{kanigel_protected_2007,
	title = {Protected Nodes and the Collapse of Fermi Arcs in High-$T_c$ Cuprate Superconductors},
	author = {Kanigel, A. and Chatterjee, U. and Randeria, M. and Norman, M. R. and Souma, S. and Shi, M. and Li, Z. Z. and Raffy, H. and Campuzano, J. C.},
	journal = {Phys. Rev. Lett.},
	volume = {99},
	pages = {157001},
	year = {2007},
	doi = {10.1103/PhysRevLett.99.157001}
}

@article{wu_fermi_2007,
	title = {Fermi liquid instabilities in the spin channel},
	author = {Wu, Congjun and Sun, Kai and Fradkin, Eduardo and Zhang, Shou-Cheng},
	journal = {Phys. Rev. B},
	volume = {75},
	pages = {115103},
	year = {2007},
	doi = {10.1103/PhysRevB.75.115103}
}

@article{sarkar_spin-split_2025,
	title = {Spin-split magnon bands induce pure spin current in insulating altermagnets},
	author = {Sarkar, Sankar and Agarwal, Amit},
	journal = {Phys. Rev. B},
	volume = {112},
	pages = {195420},
	year = {2025},
	doi = {10.1103/6g22-7s97}
}

@article{rostami_fermi_2025,
	title = {Fermi liquid theory of $d$-wave altermagnets: demon modes and {Fano}-demon states},
	author = {Rostami, Habib and Hofmann, Johannes},
	journal = {Phys. Rev. Lett.},
	volume = {135},
	pages = {236701},
	year = {2025},
	doi = {10.1103/zqz6-sq2n}
}

@article{das_realizing_2024,
	title = {Realizing Altermagnetism in {Fermi}-{Hubbard} Models with Ultracold Atoms},
	author = {Das, Purnendu and Leeb, Valentin and Knolle, Johannes and Knap, Michael},
	journal = {Phys. Rev. Lett.},
	volume = {132},
	pages = {263402},
	year = {2024},
	doi = {10.1103/PhysRevLett.132.263402}
}

@article{arovas_hubbard_2022,
	title = {The {Hubbard} Model},
	author = {Arovas, Daniel P. and Berg, Erez and Kivelson, Steven A. and Raghu, Srinivas},
	journal = {Annu. Rev. Condens. Matter Phys.},
	volume = {13},
	pages = {239},
	year = {2022},
	doi = {10.1146/annurev-conmatphys-031620-102024}
}

@article{leeb_spontaneous_2024,
	title = {Spontaneous Formation of Altermagnetism from Orbital Ordering},
	author = {Leeb, Valentin and Mook, Alexander and {\v S}mejkal, Libor and Knolle, Johannes},
	journal = {Phys. Rev. Lett.},
	volume = {132},
	pages = {236701},
	year = {2024},
	doi = {10.1103/PhysRevLett.132.236701}
}

@misc{dong_spontaneous_2025,
	title = {Spontaneous emergence of altermagnetism in the single-orbital extended {Hubbard} model},
	author = {Dong, Jin-Wei and Lin, Yu-Han and Fu, Ruiqing and Wu, Xianxin and Su, Gang and Wang, Ziqiang and Zhou, Sen},
	year = {2025},
	howpublished = {arXiv:2507.00837 [cond-mat.str-el]},
	doi = {10.48550/arXiv.2507.00837}
}

@misc{li_enhancement_2025,
	title = {Enhancement of $d$-wave Pairing in Strongly Correlated Altermagnet},
	author = {Li, Jianyu and Liu, Ji and Yang, Xiaosen and Tang, Ho-Kin},
	year = {2025},
	howpublished = {arXiv:2505.12342 [cond-mat.supr-con]},
	doi = {10.48550/arXiv.2505.12342}
}

@article{vojta_lattice_2009,
	title = {Lattice symmetry breaking in cuprate superconductors: stripes, nematics, and superconductivity},
	author = {Vojta, Matthias},
	journal = {Adv. Phys.},
	volume = {58},
	pages = {699},
	year = {2009},
	doi = {10.1080/00018730903122242}
}

@article{ghorashi_altermagnetic_2024,
	title = {Altermagnetic Routes to {Majorana} Modes in Zero Net Magnetization},
	author = {Ghorashi, Sayed Ali Akbar and Hughes, Taylor L. and Cano, Jennifer},
	journal = {Phys. Rev. Lett.},
	volume = {133},
	pages = {106601},
	year = {2024},
	doi = {10.1103/PhysRevLett.133.106601}
}

@article{smejkal_emerging_2022,
	title = {Emerging Research Landscape of Altermagnetism},
	author = {{\v S}mejkal, Libor and Sinova, Jairo and Jungwirth, Tomas},
	journal = {Phys. Rev. X},
	volume = {12},
	pages = {040501},
	year = {2022},
	doi = {10.1103/PhysRevX.12.040501}
}

@article{neehus_projectively_2025,
	title = {Projectively Implemented Altermagnetism in an Exactly Solvable Quantum Spin Liquid},
	author = {Neehus, Avedis and Rosch, Achim and Knolle, Johannes and Seifert, Urban F. P.},
	journal = {Phys. Rev. Lett.},
	volume = {135},
	pages = {256504},
	year = {2025},
	doi = {10.1103/cnk8-vnxg}
}

@article{li_exploring_2025,
	title = {Exploring $d$-wave magnetism in cuprates from oxygen moments},
	author = {Li, Ying and Leeb, Valentin and Wohlfeld, Krzysztof and Valent{\'i}, Roser and Knolle, Johannes},
	journal = {Phys. Rev. B},
	volume = {112},
	pages = {125139},
	year = {2025},
	doi = {10.1103/PhysRevB.112.125139}
}

@article{sim_pair_2025,
	title = {Pair density waves and supercurrent diode effect in altermagnets},
	author = {Sim, GiBaik and Knolle, Johannes},
	journal = {Phys. Rev. B},
	volume = {112},
	pages = {L020502},
	year = {2025},
	doi = {10.1103/b7rh-v7nq}
}

@article{macdonald_t_1988,
	title = {$t/U$ expansion for the {Hubbard} model},
	author = {MacDonald, A. H. and Girvin, S. M. and Yoshioka, D.},
	journal = {Phys. Rev. B},
	volume = {37},
	pages = {9753},
	year = {1988},
	doi = {10.1103/PhysRevB.37.9753}
}

@misc{hegde_charge_2024,
	title = {Charge asymmetry in the {Heisenberg} model},
	author = {Hegde, Rohit},
	year = {2024},
	howpublished = {arXiv:2412.07013 [cond-mat.mes-hall]},
	doi = {10.48550/arXiv.2412.07013}
}

@article{wu_pair_2023,
	title = {Pair Density Wave Order from Electron Repulsion},
	author = {Wu, Yi-Ming and Nosov, P. A. and Patel, Aavishkar A. and Raghu, S.},
	journal = {Phys. Rev. Lett.},
	volume = {130},
	pages = {026001},
	year = {2023},
	doi = {10.1103/PhysRevLett.130.026001}
}

@article{christos_model_2023,
	title = {A model of $d$-wave superconductivity, antiferromagnetism, and charge order on the square lattice},
	author = {Christos, Maine and Luo, Zhu-Xi and Shackleton, Henry and Zhang, Ya-Hui and Scheurer, Mathias S. and Sachdev, Subir},
	journal = {Proc. Natl. Acad. Sci. U.S.A.},
	volume = {120},
	number = {21},
	pages = {e2302701120},
	year = {2023},
	doi = {10.1073/pnas.2302701120}
}

@article{samajdar_polaronic_2024,
	title = {Polaronic mechanism of Nagaoka ferromagnetism in {Hubbard} models},
	author = {Samajdar, Rhine and Bhatt, R. N.},
	journal = {Phys. Rev. B},
	volume = {109},
	pages = {235128},
	year = {2024},
	doi = {10.1103/PhysRevB.109.235128}
}

@article{timusk_pseudogap_1999,
	title = {The pseudogap in high-temperature superconductors: an experimental survey},
	author = {Timusk, Tom and Statt, Bryan},
	journal = {Rep. Prog. Phys.},
	volume = {62},
	pages = {61--122},
	year = {1999},
	doi = {10.1088/0034-4885/62/1/002}
}

@article{norman_pseudogap_2005,
	title = {The pseudogap: friend or foe of high $T_c$?},
	author = {Norman, M. R. and Pines, D. and Kallin, C.},
	journal = {Adv. Phys.},
	volume = {54},
	number = {8},
	pages = {715--733},
	year = {2005},
	doi = {10.1080/00018730500459906}
}

@article{loeser_excitation_1996,
  author  = {A. G. Loeser and others},
  title   = {Excitation Gap in the Normal State of Underdoped {Bi$_{2}$Sr$_{2}$CaCu$_{2}$O$_{8+\delta}$}},
  journal = {Science},
  volume  = {273},
  pages   = {325--329},
  year    = {1996},
  doi     = {10.1126/science.273.5273.325}
}

@article{ding_spectroscopic_1996,
  author  = {H. Ding and others},
  title   = {Spectroscopic evidence for a pseudogap in the normal state of underdoped high-{$T_c$} superconductors},
  journal = {Nature},
  volume  = {382},
  pages   = {51--54},
  year    = {1996},
  doi     = {10.1038/382051a0}
}

@article{norman_destruction_1998,
  author  = {M. R. Norman and others},
  title   = {Destruction of the Fermi surface in underdoped high-{$T_c$} superconductors},
  journal = {Nature},
  volume  = {392},
  pages   = {157--160},
  year    = {1998},
  doi     = {10.1038/32366}
}

@article{howald_periodic_2003,
  author  = {C. Howald and others},
  title   = {Periodic density-of-states modulations in superconducting {Bi$_{2}$Sr$_{2}$CaCu$_{2}$O$_{8+\delta}$}},
  journal = {Phys. Rev. B},
  volume  = {67},
  pages   = {014533},
  year    = {2003},
  doi     = {10.1103/PhysRevB.67.014533}
}

@article{vershinin_local_2004,
  author  = {M. Vershinin and others},
  title   = {Local Ordering in the Pseudogap State of the High-{$T_c$} Superconductor {Bi$_{2}$Sr$_{2}$CaCu$_{2}$O$_{8+\delta}$}},
  journal = {Science},
  volume  = {303},
  pages   = {1995--1998},
  year    = {2004},
  doi     = {10.1126/science.1093384}
}

@article{alloul_89y_1989,
  author  = {H. Alloul and T. Ohno and P. Mendels},
  title   = {{$^{89}$Y} NMR evidence for a fermi-liquid behavior in {YBa$_{2}$Cu$_{3}$O$_{6+x}$}},
  journal = {Phys. Rev. Lett.},
  volume  = {63},
  pages   = {1700--1703},
  year    = {1989},
  doi     = {10.1103/PhysRevLett.63.1700}
}

@article{warren_cu_1989,
  author  = {W. W. Warren Jr. and others},
  title   = {Cu spin dynamics and superconducting precursor effects in planes above {$T_c$} in {YBa$_{2}$Cu$_{3}$O$_{6.7}$}},
  journal = {Phys. Rev. Lett.},
  volume  = {62},
  pages   = {1193--1196},
  year    = {1989},
  doi     = {10.1103/PhysRevLett.62.1193}
}

@article{ando_electronic_2004,
  author  = {Y. Ando and others},
  title   = {Electronic phase diagram of high-{$T_c$} cuprate superconductors from a mapping of the in-plane resistivity curvature},
  journal = {Phys. Rev. Lett.},
  volume  = {93},
  pages   = {267001},
  year    = {2004},
  doi     = {10.1103/PhysRevLett.93.267001}
}

@article{emery_phase_1995,
  author  = {V. J. Emery and S. A. Kivelson},
  title   = {Importance of phase fluctuations in superconductors with small superfluid density},
  journal = {Nature},
  volume  = {374},
  pages   = {434--437},
  year    = {1995},
  doi     = {10.1038/374434a0}
}

@article{chakravarty_hidden_2001,
  author  = {S. Chakravarty and others},
  title   = {Hidden order in the cuprates},
  journal = {Phys. Rev. B},
  volume  = {63},
  pages   = {094503},
  year    = {2001},
  doi     = {10.1103/PhysRevB.63.094503}
}

@article{varma_nonfermi_1997,
  author  = {C. M. Varma},
  title   = {Non-Fermi-liquid states and pairing instability of a general model of copper oxide metals},
  journal = {Phys. Rev. B},
  volume  = {55},
  pages   = {14554--14580},
  year    = {1997},
  doi     = {10.1103/PhysRevB.55.14554}
}

@article{wenlee_underdoped_1996,
  author  = {X.-G. Wen and P. A. Lee},
  title   = {Theory of underdoped cuprates},
  journal = {Phys. Rev. Lett.},
  volume  = {76},
  pages   = {503--506},
  year    = {1996},
  doi     = {10.1103/PhysRevLett.76.503}
}

@article{lee_nagaosa_wen_2006,
  author  = {P. A. Lee and N. Nagaosa and X.-G. Wen},
  title   = {Doping a Mott insulator: Physics of high-temperature superconductivity},
  journal = {Rev. Mod. Phys.},
  volume  = {78},
  pages   = {17--85},
  year    = {2006},
  doi     = {10.1103/RevModPhys.78.17}
}

@article{yang_phenomenological_2006,
  author  = {K.-Y. Yang and T. M. Rice and F.-C. Zhang},
  title   = {Phenomenological theory of the pseudogap state},
  journal = {Phys. Rev. B},
  volume  = {73},
  pages   = {174501},
  year    = {2006},
  doi     = {10.1103/PhysRevB.73.174501}
}

@article{read_sachdev_valence_1989,
  author  = {N. Read and S. Sachdev},
  title   = {Valence-bond and spin-Peierls ground states of low-dimensional quantum antiferromagnets},
  journal = {Phys. Rev. Lett.},
  volume  = {62},
  pages   = {1694--1697},
  year    = {1989},
  doi     = {10.1103/PhysRevLett.62.1694}
}

@article{sachdev_order_2003,
  author  = {S. Sachdev},
  title   = {Colloquium: Order and quantum phase transitions in the cuprate superconductors},
  journal = {Rev. Mod. Phys.},
  volume  = {75},
  pages   = {913--932},
  year    = {2003},
  doi     = {10.1103/RevModPhys.75.913}
}

\end{document}